# Area Optimized Quasi Delay Insensitive Majority Voter for TMR Applications

P. Balasubramanian, D.L. Maskell
School of Computer Science and Engineering
Nanyang Technological University
Singapore 639798
{balasubramanian, asdouglas}@ntu.edu.sg

N.E. Mastorakis
Department of Industrial Engineering
Technical University of Sofia
Sofia 1000, Bulgaria
mastor@tu-sofia.bg

*Abstract*—Mission-critical and safety-critical applications generally tend to incorporate triple modular redundancy (TMR) to embed fault tolerance in their physical implementations. In a TMR realization, an original function block, which may be a circuit or a system, and two exact copies of the function block are used to successfully overcome any temporary fault or permanent failure of an arbitrary function block during the routine operation. The corresponding outputs of the function blocks are majority voted using 3-input majority voters whose outputs define the outputs of a TMR realization. Hence, a 3-input majority voter forms an important component of a TMR realization. Many synchronous majority voters and an asynchronous non-delay insensitive majority voter have been presented in the literature. Recently, quasi delay insensitive (QDI) asynchronous majority voters for TMR applications were also discussed in the literature. In this regard, this paper presents a new QDI asynchronous majority voter for TMR applications, which is better optimized in area compared to the existing QDI majority voters. The proposed QDI majority voter requires 30.2% less area compared to the best of the existing QDI majority voters, and this could be useful for resource-constrained fault tolerance applications. The example QDI TMR circuits were implemented using a 32/28nm complementary metal oxide semiconductor (CMOS) process. The delay insensitive dual rail code was used for data encoding, and 4-phase return-to-zero and return-to-one handshake protocols were used for data communication.

## I. Introduction

Mission-critical applications such as space and aerospace systems, and safety-critical applications such as electrical power transmission and distribution systems, nuclear power plants, and banking and financial systems etc. generally incorporate some form of N-modular redundancy (NMR) in their implementations to embed fault tolerance [1,2]. Fault tolerance is important in mission-critical and safety-critical applications to avoid a sudden catastrophic failure and a consequent system shutdown.

n NMR, N identical function blocks are used and the correct operation of (N+1)/2 function blocks is necessary. With this arrangement, an NMR is able to tolerate the temporary faults or permanent failures of maximum of (N–1)/2 function blocks. Triple modular redundancy (TMR or 3MR) is the basic version of NMR in which 3 identical function blocks are used and the fault or failure of any one of the function blocks is tolerated. The corresponding outputs of the function blocks in an NMR implementation are combined using N-input majority voters, which produce the NMR outputs based on the Boolean majority. The number of majority voters to be used depends on the number of outputs produced by a function block. For example, if a function block produces K outputs, then K N-input majority voters are used [3]; for a TMR, K 3-input majority voters would be used. A 3-input majority voter is an important component in a TMR realization. Besides this, the 3-input majority voter also finds use in other redundancy schemes such as majority and minority voting based redundancy [4–7] and self-healing redundancy [8] schemes, which are suitable for mission-critical and safety-critical applications.

In the literature, designs of many majority voters for TMR applications have been described based on the synchronous design style [9–13]. Also, the design of an asynchronous majority voter for an asynchronous TMR realization was presented in [14]. However, the asynchronous majority voter of [14] is not delay insensitive since it corresponds to a bundled data handshake protocol, and it is not robust. Recently, quasi delay insensitive (QDI) asynchronous designs of a 3-input majority voter based on different QDI logic synthesis methods were explored and discussed in [15]. Reference [15] assumes significance in view of the fact that the QDI design [16,17] is an efficient, practical and robust alternative to synchronous design due to the following advantages: modular, low power, tolerant to process, voltage and temperature (PVT) variations, less susceptible to electromagnetic interference, resistant to side channel attacks [18], and self-checking [19].

Reference [15] has considered the designs of QDI majority voters for TMR applications, based on some strong indication QDI logic synthesis methods, and identified which QDI majority voter is of low power and high speed, and which QDI majority voter is suitable for less area. In this regard, this paper proposes a new QDI majority voter suitable for TMR applications, which features a reduced area compared to the QDI majority voters discussed in [15].

The remainder of this paper is organized as follows. The basics of QDI circuit design are discussed in Section 2. The topology of a QDI TMR realization is illustrated in Section 3, and the novel design of the QDI majority voter is presented. Section 4 gives the simulation results corresponding to various QDI majority voters including the proposed voter for QDI TMR implementations and makes a comparative analysis. Finally, the conclusions are provided in Section 5.

This work is supported by an Academic Research Fund Tier-2 research award of the Ministry of Education, Singapore under grant MOE2018-T2-2-024.

## II. QDI Circuit Design – Basics

The basic principles of QDI circuit design are discussed under two sub-headings viz. delay insensitive data encoding and handshaking, and strong indication.

### A. Data Insensitive Encoding and 4-Phase Handshaking

The block schematic of a QDI circuit stage is shown in the middle of Fig. 1 and its correlation with the sender (SR)–receiver (RR) analogy is also shown. A QDI circuit stage consists of the current stage register bank (CSRB) and the next stage register bank (NSRB), a QDI circuit sandwiched between these register banks, completion detectors that are associated with the register banks, and acknowledgement input (ACKIT) and acknowledgement output (ACKOT) signals which are exchanged between the register banks. ACKIT and ACKOT are exchanged by following a 4-phase handshake protocol, which may involve either return-to-zero (RTZ) [20] signalling or return-to-one (RTO) [21] signalling.

In Fig. 1, the CSRB and NSRB are analogous to SR and RR respectively. CSRB and NSRB comprise a number of registers with one register allotted for each rail of an encoded data. CSRB may be interpreted as the input register bank while NSRB may be interpreted as the output register bank. In fact, NSRB serves as the CSRB for the next stage of the QDI circuit. Here, a register implies a 2-input Muller C-element [22]. The C-element will output binary 1 if all its inputs are binary 1, and will output binary 0 if all its inputs are binary 0. If the inputs to a C-element are not identical then the C-element would retain its existing steady state. The circles with the marking 'C' represent the C-elements in the figures. The symbol, output equation, and gate level and transistor level realizations of a 2-input C-element are shown in Fig. 2.

The inputs and outputs of a QDI circuit are encoded using a delay insensitive code [23]. Among the family of delay insensitive codes, the dual rail code is widely used for QDI circuit designs. In Fig. 1, (J1, J0) and (K1, K0) represent the delay insensitive dual rail encoded inputs of the single rail inputs J and K respectively. According to dual rail encoding and RTZ handshaking, an input J is encoded as (J1, J0) where J = 1 is given by J1 = 1 and J0 = 0, and J = 0 is given by J0 = 1 and J1 = 0. These assignments are called 'data'. The assignment J1 = J0 = 0 is called the 'spacer', and the assignment J1 = J0 = 1 is considered to be indeterminate to maintain the delay insensitivity.

On the other hand, according to dual rail encoding and RTO handshaking, an input J is encoded as (J1, J0) where J = 1 is given by J1 = 0 and J0 = 1, and J = 0 is given by J0 = 0 and J1 = 1. These assignments are called data. The assignment J1 = J0 = 1 is called the spacer, and the assignment J1 = J0 = 0 is deemed indeterminate to maintain the delay insensitivity.

Four steps are involved in RTZ and RTO handshaking, and hence they are referred to as 4-phase handshake protocols. The RTZ and RTO handshake signalling are described below by referring to Fig. 1. Handshaking is performed between CSRB and NSRB while involving the QDI circuit.

According to RTZ handshaking, first, the dual rail data bus comprising (J1, J0) and (K1, K0) initially assumes the spacer and ACKIT = 1. After the CSRB sends data, rising signal transitions i.e. 0 to 1 will occur on one of the rails of each dual rail encoded input of the data bus. Second, the NSRB, after receiving the processed data from the QDI circuit, would drive ACKOT to 1. Third, the CSRB would wait for ACKIT to become 0, and after this happens, the data bus would once again assume the spacer. Finally, i.e. fourth, after a finite and positive unbounded time duration elapses, the NSRB would receive the spacer from the QDI circuit and would drive ACKOT to 0. Consequently, ACKIT will become 1. With this process, one data transaction is said to be complete and the next data transaction may commence. As per RTZ handshaking, the inputs are supplied following the sequence of data-spacer-data-spacer, and so forth.

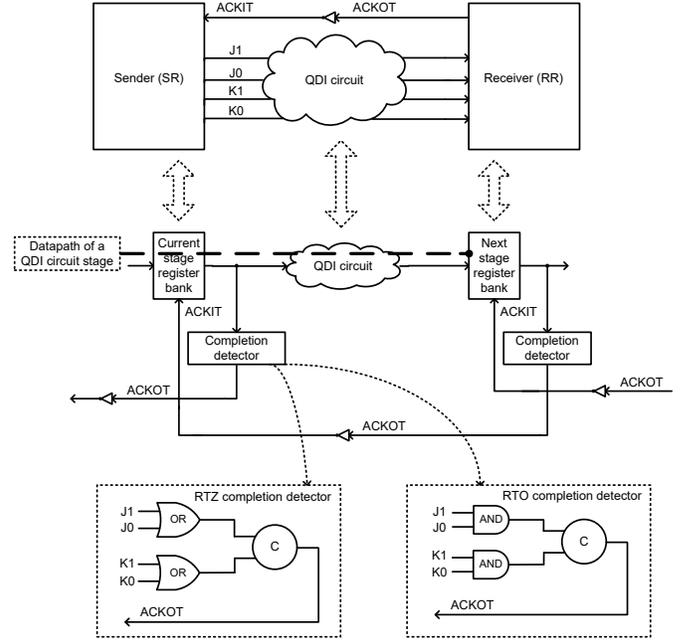

Fig. 1. A typical quasi delay insensitive (QDI) circuit stage. Return-to-zero (RTZ) and return-to-one (RTO) completion detectors for the example dual-rail inputs (J1, J0) and (K1, K0) are shown enclosed within the dotted rectangles. The critical datapath is highlighted by the thick dashed line, which involves the current stage register bank and the QDI circuit.

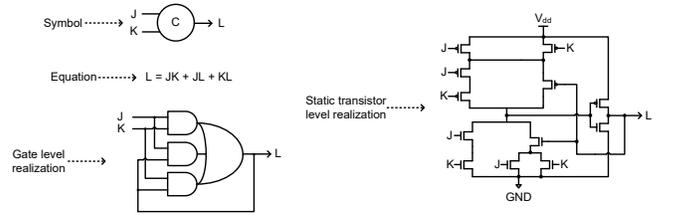

Fig. 2. Description of a 2-input Muller C-element with J and K as the inputs and L as the output.

According to RTO handshaking, first, ACKIT = 1. The CSRB now sends the spacer i.e. all ones, and as a result rising signal transitions will occur on all the rails of the data bus. Second, the NSRB, after receiving the spacer from the QDI circuit would drive ACKOT to 1. Third, the CSRB would wait for ACKIT to become 0, and after this happens, the data bus would send the data by permitting falling signal transitions i.e.

1 to 0 to occur on one of the rails of each dual rail encoded input of the data bus. Finally, i.e. fourth, after a finite and positive unbounded time duration elapses, the NSRB would receive the processed data from the QDI circuit and subsequently drive ACKOT to 0. Consequently, ACKIT will become 1. With this process, one data transaction is said to be complete and the next data transaction may commence. As per RTO handshaking, the inputs are supplied following the sequence of spacer-data-spacer-data, and so forth.

Forward latency, reverse latency, and cycle time are the timing parameters of interest in a QDI circuit. Forward latency refers to the worst-case propagation delay encountered in the data path (i.e. the thick dashed line shown in Fig. 1) for processing the data, and reverse latency refers to the worst-case propagation delay encountered in the data path for processing the spacer. From the above discussion, it may be noted that a data transaction in a QDI circuit would comprise the forward latency as well as the reverse latency. The summation of forward and reverse latencies gives the cycle time. The cycle time governs the speed of a QDI circuit, which is equivalent to the clock period of a synchronous digital logic circuit.

The gate level detail of the example completion detectors corresponding to 4-phase RTZ and RTO handshaking is shown at the bottom of Fig. 1 within the dotted rectangles. The completion detector indicates i.e. acknowledges the receipt of all the primary inputs given to a QDI circuit. In the case of RTZ handshaking, ACKOT is provided by employing a 2-input OR gate to respectively combine the dual rails of each encoded input, and the outputs of all such 2-input OR gates are then synchronized using a C-element or a tree of C-elements. In the case of RTO handshaking, ACKOT is provided by employing a 2-input AND gate to combine the respective dual rails of each encoded input, and the outputs of all such 2-input AND gates are then synchronized using a C-element or a tree of C-elements. ACKIT is the Boolean complement of ACKOT and vice-versa.

*B. Strong Indication*

QDI circuits are generally classified as strong indication [24,25], weak indication [24,26] and early output [27] types. A strong indication circuit would commence the processing of data or spacer after receiving all the data or spacer. In other words, if the complete set of input data or spacer is not supplied, a strong indication circuit will not process the data or spacer. The input/output timing behaviour of a strong indication (QDI) circuit is illustrated via Fig. 3a which corresponds to RTZ handshaking, and Fig. 3b which corresponds to RTO handshaking.

It is observed in [15] that since the majority voter has only one dual rail primary output, a strongly indicating realization is necessary and weak indication or early output realizations are not suitable. This is due to two reasons. Firstly, weak indication and early output QDI circuits are able to produce a subset or all of the primary outputs after receiving a subset of the primary inputs, and a weak indication circuit requires at least a pair of dual rail primary outputs. Secondly, any late arrival of an input to a weak indication or an early output majority voter could result in a gate orphan. A gate orphan is a signal transition i.e. 0 to 1 or 1 to 0 that occurs in an intermediate gate output of a QDI circuit when it is not acknowledged by the output of a gate present in the subsequent logic level. Gate orphans are likely to affect the robustness of QDI circuits [27,28], and they are best avoided. The gate orphan has been explained through some example scenarios in [29–32], and an interested reader may refer to the same for details. A QDI circuit is not only mandated to produce the correct primary output(s) for the given primary input(s) but also required to properly indicate the completion of internal processing of data and spacer within the circuit [20].

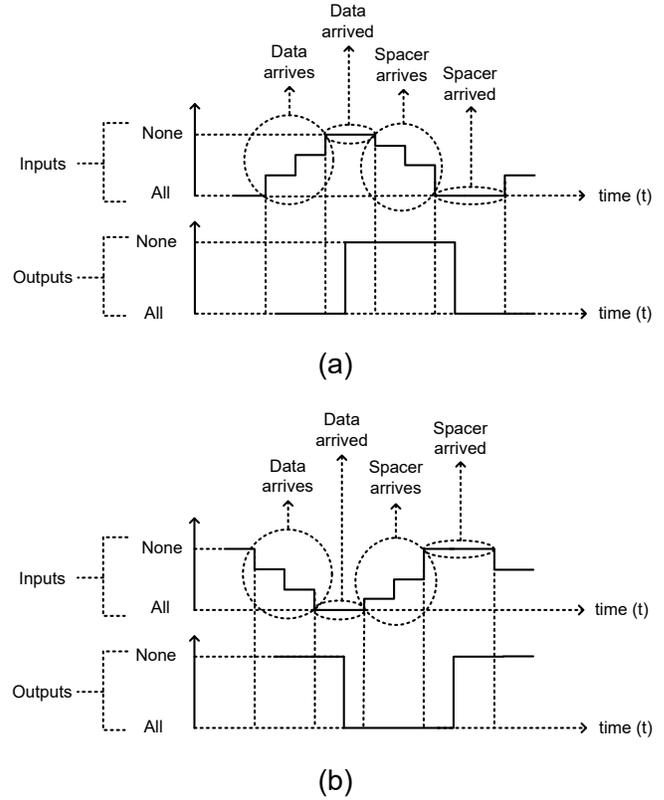

Fig. 3. Input/output timing behavior of a strong indication circuit corresponding to: (a) RTZ handshaking; and (b) RTO handshaking.

## III. QDI TMR Realization and the Proposed QDI 3-Input Majority Voter

The topology of a QDI TMR circuit is shown in Fig. 4, which consists of three identical function blocks. Similar to [15], we considered using an early output QDI full adder [33] for the function blocks. The full adder [34] adds three input bits and produces two output bits, namely the sum and any carry overflow. It is noted in [35] that the early output QDI full adder of [33], when replicated and cascaded, results in an efficient QDI ripple carry adder. In Fig. 4, (J1, J0), (K1, K0) and (L1, L0) represent the dual rail primary inputs to the function blocks, and (P1, P0), (Q1, Q0) and (R1, R0) represent the dual rail primary outputs of the function blocks, which are in turn given as inputs to the QDI majority voter. The majority voter performs a voting based on the Boolean majority and produces the primary output (M1, M0), which represents the output of the TMR realization. The dual rail output expressions of the majority voter are given by (1) and (2) [15]. Equations (1) and (2) signify the correct operation of any 2 function blocks in the TMR. The correct operation of all the 3 function blocks i.e. P1Q1R1 with respect to (1) and P0Q0R0 with respect to (2) are implicitly covered by (1) and (2), as noted in [15].

$$M1 = P1Q1 + Q1R1 + P1R1 \qquad (1)$$

$$M0 = P0Q0 + Q0R0 + P0R0 \qquad (2)$$

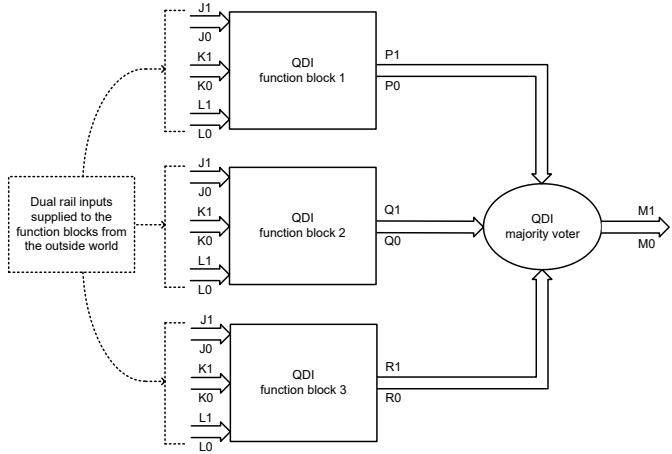

Fig. 4. Block diagram of an example QDI triple modular redundancy (TMR) realization consisting of 3 identical function blocks each producing a dual rail output which are combined using a majority voter. Same dual rail inputs are supplied to all the function blocks from the outside world.

It has been discussed in [15] how strong indication majority voters corresponding to different strong indication logic synthesis methods [36–39] were implemented conforming to RTZ and RTO handshaking. An interested reader is referred to [15] for the corresponding circuit realizations and the repetition is avoided here for brevity.

The proposed strong indication majority voter is shown in Fig. 5. Fig. 5a shows the majority voter that corresponds to RTZ handshaking, and Fig. 5b shows the majority voter that corresponds to RTO handshaking. The logic rules governing the conversion of a QDI circuit corresponding to RTZ handshaking into one that corresponds to RTO handshaking and vice-versa have been stated [40] and proved [41]. Excepting the C-elements and their respective inputs, if the remaining logic gates in a QDI circuit are replaced by their corresponding duals, a QDI circuit belonging to one handshake protocol (RTZ or RTO) would be transformed into a QDI circuit belonging to another handshake protocol (RTO or RTZ). Assuming A, B and C are the inputs, the AO222 and OA222 gates shown in Figs. 5a and 5b would implement (AB + BC + AC) and [(A+B) (B+C) (A+C)] respectively, after logic factoring [42].

It is described in [15] how a naïve implementation of equations (1) and (2) could result in the problem of gate orphans. Therefore, (1) and (2) are synthesized using two complex gates, namely AO222 gates, as shown in Fig. 5a. The dual of the AO222 gate is the complex gate OA222, which is used to implement the majority voter that corresponds to RTO handshaking, as shown in Fig. 5b. Because complex gates are used to synthesize V1 and V0, it is not required that the sum of product expressions (1) and (2) should be transformed into disjoint sum of product expressions [43], which are commonly used in QDI circuit designs.

The use of complex gates alone would not suffice to make the proposed majority voter strongly indicating. For example, referring to Fig. 5a, let us assume that after an RTZ phase, P1 = Q1 = 1, and that R1 is also 1 but R1 arrives lately. In this case, one of the AO222 gates will output 1 and thus the internal output NM1 would assume 1 regardless of the late arrival of R1, leading to an early output majority voter.

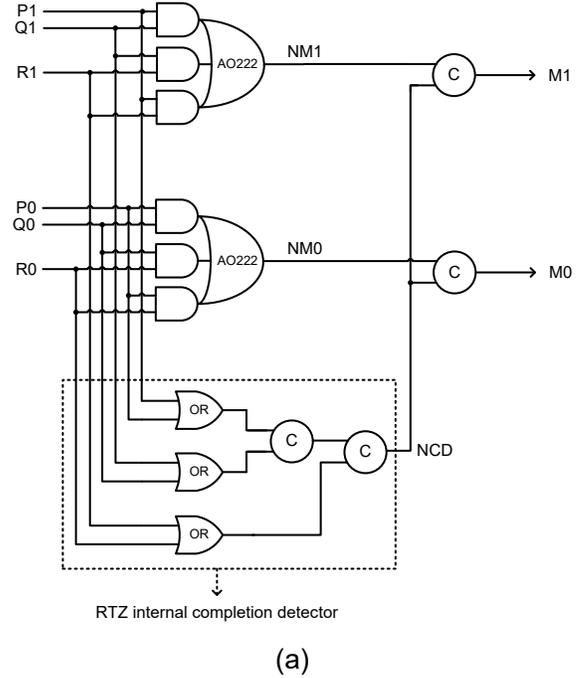

(a)

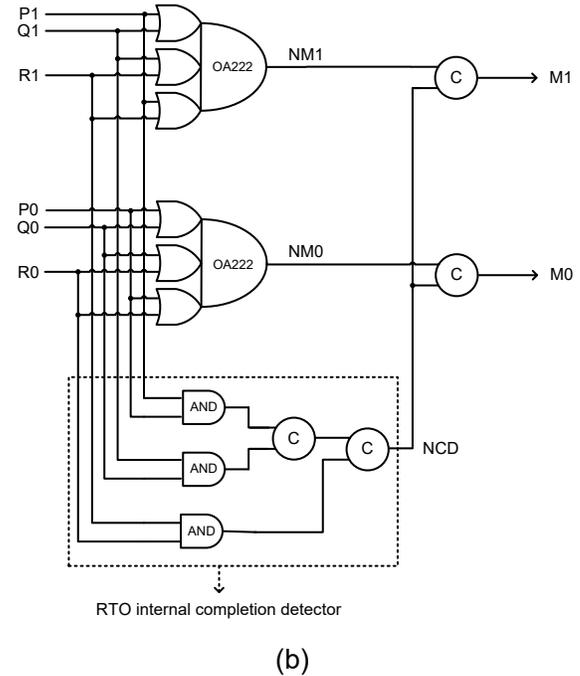

(b)

Fig. 5. Proposed strong indication 3-input majority voter, corresponding to: (a) RTZ handshaking; and (b) RTO handshaking.

Supposing the internal output NM1 has been used to represent the voter output M1 then the late arrival of R1 would be termed as a gate orphan. Similarly, if we assume that after an RTZ phase, P0 = Q0 = 1, and that R0 is also 1 but R0 arrives lately. Again, one of the AO222 gates will output 1 and the internal output NM0 would assume 1 regardless of the late arrival of R0, again resulting in an early output majority voter. Supposing the internal output NM0 has been used to represent the voter output M0, the late arrival of R1 would be termed as a gate orphan. In this context, it may be noted that considering and analysing the late arrival of signals is necessary in a QDI circuit to ensure that delay insensitivity would be guaranteed. To eliminate the problem of gate orphans, an internal completion detector is provided, which is shown within the dotted rectangles in Figs. 5a and 5b. Although (NM1, NM0) is logically equivalent to (M1, M0), the output of the internal completion detector i.e. NCD is synchronized with (NM1, NM0) using two 2-input C-elements to yield the majority voter's primary output (M1, M0), as shown in Figs. 5a and 5b.

The internal completion detectors, shown in Figs. 5a and 5b, are similar in logic to those shown in Fig. 1 with respect to RTZ and RTO handshaking. In Fig. 1, the completion detector is associated with the entire QDI circuit, which in our case includes the function blocks and the voters as depicted in Fig. 4. On the other hand, the internal completion detectors shown in Figs. 5a and 5b just form a part of the proposed QDI majority voter. The internal completion detectors mainly help to enforce the strong indication property. In Figs. 5a and 5b, the internal completion detectors will strongly indicate the entire arrival of data as well as spacer by producing 1 and 0 respectively for RTZ handshaking, and 0 and 1 respectively for RTO handshaking. Also, the signal transitions would occur monotonically in the majority voter from the primary inputs up to the primary output [44], i.e. rising or falling signal transitions on the primary inputs would be uniformly propagated throughout the entire circuit up to the primary outputs.

## IV. SIMULATION RESULTS

Example QDI TMR circuits were implemented as a typical QDI circuit stage comprising an input register bank and the TMR circuit. The TMR circuits employed an early output QDI full adder [33] for the function blocks and used the strong indication majority voters discussed in [15] and also the proposed strong indication majority voter. Each TMR implementation comprised 3 full adders and 2 majority voters with one majority voter assigned for the sum output and another for the carry output.

The QDI TMR circuits were implemented using a 32/28nm CMOS process [45]. Excepting the 2-input C-element which was custom-realized as a static complex gate implementation, as shown in Fig. 2, all the other simple and complex gates used to construct the TMR circuits were utilized directly from the standard digital cell library [45]. A typical case PVT specification of the standard digital cell library was considered for the simulations by using a supply voltage of 1.05V under an operating junction temperature of 25ºC. A virtual clock was used just superficially to constraint the input and output ports of the QDI TMR circuits to estimate the critical path delay (forward latency). However, the clock did not form a part of the designs.

The same test benches, as used in [15], were used for this work to verify the functionality and to estimate the switching activity, which would pave the way for a straightforward simulation-based comparison. The switching activity data obtained from the functional simulations were used to estimate the average power dissipation of the QDI TMR circuits. The functional simulations covered all possible scenarios for the function blocks and it was confirmed that majority voted outputs are produced.

Screenshots of portions of the simulation waveforms corresponding to QDI TMR circuits, which employ the proposed majority voter CG_MV, are shown in Figs. 6 and 7, which correspond to RTZ and RTO handshaking respectively. In Figs. 6 and 7, on the left-side, (SUM01,SUM00), (SUM11,SUM10) and (SUM21,SUM20), given within the purple boxes, represent the dual rail sum outputs while (CARRYOUT01,CARRYOUT00), (CARRYOUT11,CARRYOUT10) and (CARRYOUT21,CARRYOUT20), given within the orange boxes, represent the dual rail carry outputs of the 3 QDI full adders (i.e. function blocks). (SUM31,SUM30) and (CARRYOUT31,CARRYOUT30), highlighted in blue on the left-side, represent the majority voted sum and carry outputs of the QDI TMR circuits. Some distinct combinations of inputs and majority voted sum and carry outputs are highlighted within the rose and yellow rectangles in Figs. 6 and 7.

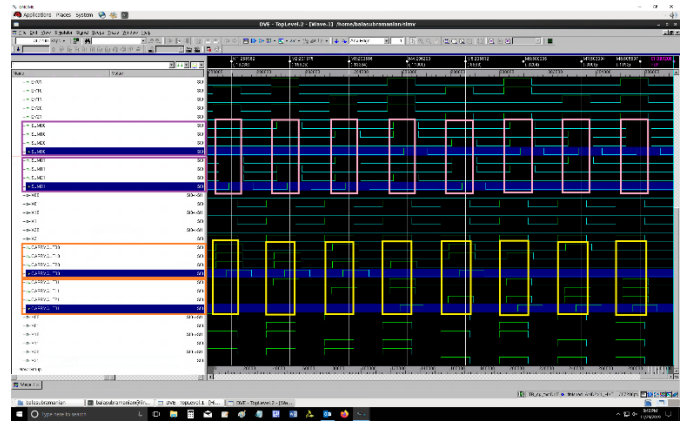

Fig. 6. Screenshot of a portion of the simulation waveforms of a QDI TMR circuit incorporating CG_MV corresponding to RTZ handshaking.

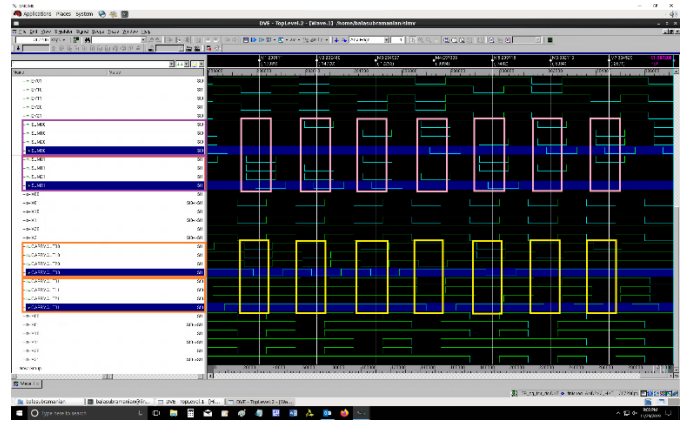

Fig. 7. Screenshot of a portion of the simulation waveforms of a QDI TMR circuit incorporating CG_MV corresponding to RTO handshaking.

Synopsys tools were used to estimate the design metrics viz. latency, area, and average power dissipation. Since strong indication majority voters are considered, the forward and reverse latencies of the QDI TMR circuits are equal. This is because strong indication circuits encounter the worst-case latency for the processing of data and spacer. Hence, the cycle time is twice that of the forward latency. The estimated design metrics are given in Table I. In Table I, the legends Singh_MV, DIMS_MV, Toms_MV and SI_MV denote the 3-input QDI majority voters based on [36], [37], [38] and [39] respectively. The legend 'CG_MV' represents the complex gates based 3-input QDI majority voter, proposed in this work. The split-up of power dissipation between majority voters and others (i.e. function blocks, registers and completion detector) is also given in Table I. Since the same function blocks, registers and completion detector are used to construct all the QDI TMR circuits corresponding to a specific handshake protocol, the differences in their design metrics are practically attributed to the differences in logic between the QDI majority voters.

TABLE I. DESIGN PARAMETERS OF VARIOUS QUASI DELAY INSENSITIVE TRIPLE MODULAR REDUNDANCY CIRCUITS EMPLOYING DIFFERENT MAJORITY VOTERS, ESTIMATED USING A 32/28-NM COMPLEMENTARY METAL OXIDE SEMICONDUCTOR TECHNOLOGY

| Majority Voter used | Cycle Time (ns) | Area ($\mu m^2$) | Power Dissipation ($\mu W$) | | |
|---|---|---|---|---|---|
| | | | Voters | Others | Total |
| *Corresponding to RTZ handshake protocol* | | | | | |
| Singh_MV | 2.32 | 267.61 | 36.80 | 196.00 | 232.8 |
| DIMS_MV | 1.78 | 259.99 | 21.65 | 197.95 | 219.6 |
| Toms_MV | 1.88 | 252.87 | 22.26 | 197.04 | 219.3 |
| SI_MV | 1.94 | 240.67 | 40.55 | 195.75 | 236.3 |
| CG_MV | 1.92 | 218.31 | 37.49 | 195.01 | 232.5 |
| *Corresponding to RTO handshake protocol* | | | | | |
| Singh_MV | 2.26 | 267.61 | 36.84 | 196.46 | 233.3 |
| DIMS_MV | 1.72 | 255.92 | 19.29 | 198.41 | 217.7 |
| Toms_MV | 1.88 | 252.87 | 22.61 | 197.49 | 220.1 |
| SI_MV | 1.92 | 240.67 | 40.36 | 196.64 | 237.0 |
| CG_MV | 1.90 | 218.31 | 37.27 | 195.03 | 232.3 |

From Table I, the QDI TMR circuits comprising DIMS_MV and Toms_MV dissipate nearly the same total power for RTZ and RTO handshaking. However, considering the power dissipation of majority voters, DIMS_MV dissipates 2.7% less power than Toms_MV with respect to RTZ handshaking and 14.7% less power with respect to RTO handshaking. The increase in power savings for DIMS_MV compared to Toms_MV with respect to RTO handshaking is due to the use of high fan-in AND gates in the case of RTO handshaking, which dissipate less power than the high fan-in OR gates used for RTZ handshaking. From Table I, it is noted that from the perspectives of power dissipation and cycle time, the QDI TMR circuit containing DIMS_MV is preferable to the rest on the basis of RTZ and RTO handshaking. The QDI TMR circuit comprising CG_MV dissipates a relatively greater power and this is due to its internal completion detector. In fact, the internal completion detectors present in CG_MV and SI_MV experience regular switching activity for the application of data and spacer and this causes an increase in their dynamic power dissipation and eventually an increase in the total power dissipation of the QDI TMR circuits employing these voters. However, the QDI TMR circuit comprising CG_MV dissipates somewhat less power than the QDI TMR circuit comprising SG_MV for both the handshake schemes. This is because despite both CG_MV and SI_MV incorporating an internal completion detector, the voter logic of CG_MV is more compact compared to the voter logic of SI_MV. This would be evident upon comparing Fig. 5a of this work with Fig. 9 of [15], and Fig. 5b of this work with Fig. 10 of [15].

The areas of some QDI TMR circuits are found to be the same for both RTZ and RTO handshaking. This is because some of the dual gates in the standard digital cell library [45] feature similar areas. For examples, minimum size 2-input AND and OR gates, and minimum size AO222 and OA222 gates in [45] respectively consume the same area. From Table I, in terms of area, the QDI TMR circuit comprising the proposed majority voter CG_MV requires less silicon compared to the other QDI TMR circuits for both the handshake protocols. This is because CG_MV requires less silicon compared to the other majority voters, as seen from Fig. 8. It is found that CG_MV requires less area than Singh_MV, DIMS_MV, Toms_MV and SI_MV by 48.7%, 44.6%, 40% and 30.2% respectively for RTZ handshaking. Based on RTO handshaking, CG_MV requires less area than Singh_MV, DIMS_MV, Toms_MV and SI_MV by 48.7%, 42.1%, 40% and 30.2% respectively.

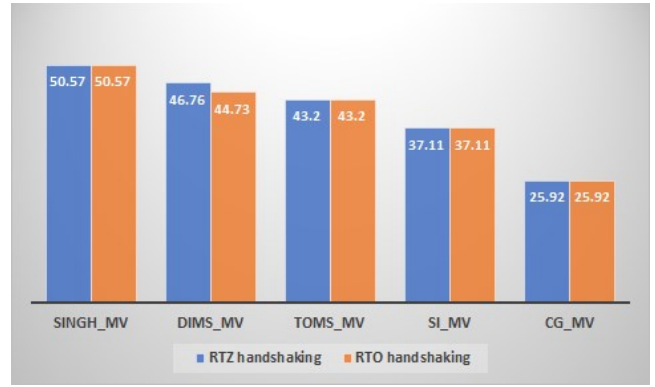

Fig. 8. Comparison of areas (in $\mu m^2$) of different QDI majority voters corresponding to both the handshake protocols.

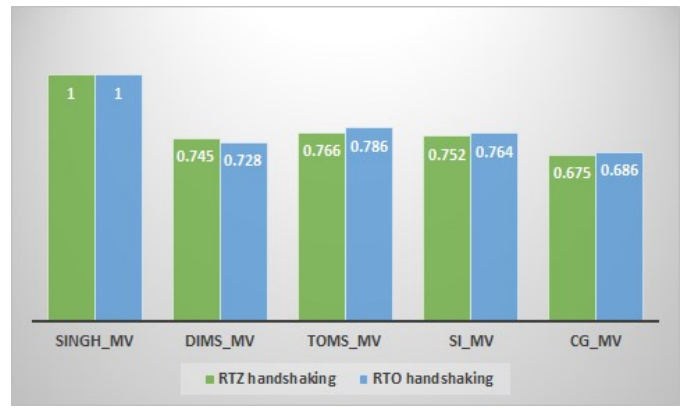

Fig. 9. Comparison of area-cycle time product (ACTP) of various QDI TMR circuits incorporating different majority voters, based on RTZ and RTO handshaking.

Quite often, a decrease in the critical path delay of a digital circuit tends to be accompanied by an increase in the area. The area-delay product (ADP) is therefore used as a notional figure of merit to interpret whether a decrease in area causes an increase in the delay. Since area and delay are desirable to be less, the ADP is also desirable to be less. With respect to a QDI design, the ADP is specified by the product of area and cycle time i.e. the area-cycle time product (ACTP). The critical path delay primarily determines the throughput in a synchronous circuit while the cycle time determines the throughput in a QDI circuit. Thus, the ACTP of a QDI circuit is equivalent to the area-delay product (ADP) of a synchronous logic circuit. Since area and cycle time are desirable to be minimum, the ACTP is also desirable to be minimum. Fig. 9 shows a comparison of the normalized area-cycle time product (ACTP) of various QDI TMR circuits corresponding to RTZ and RTO handshaking, which incorporate various majority voters. The actual ACTP of various QDI TMR circuits are first calculated from Table I. A normalization was then performed by dividing the actual ACTP of all the QDI TMR circuits with the highest ACTP corresponding to a particular handshake protocol. The least value of the normalized PCTP pertaining to a QDI TMR circuit suggests that it is the best from the ACTP perspective.

From Fig. 9, it is noted that the QDI circuit comprising CG_MV has a reduced ACTP compared to the ACTP of other QDI TMR circuits incorporating other majority voters – this observation holds good for both RTZ and RTO handshaking. With respect to RTZ handshaking, the QDI TMR circuit comprising CG_MV reports reductions in ACTP compared to the QDI TMR circuits comprising Singh_MV, DIMS_MV, Toms_MV and SI_MV by 32.5%, 9.4%, 11.9% and 10.2% respectively. For RTO handshaking, the QDI TMR circuit comprising CG_MV reports respective reductions in ACTP compared to the QDI TMR circuits comprising Singh_MV, DIMS_MV, Toms_MV and SI_MV by 31.4%, 5.8%, 12.7% and 10.2%.

V. Conclusions

This paper has presented and discussed a new robust QDI asynchronous 3-input majority voter (CG_MV), suitable for TMR realizations targeting mission-critical and safety-critical applications. Notably, the proposed CG_MV requires less area and also features a less ACTP compared to the other QDI majority voters in the existing literature. In comparison with the minimum area occupied by an existing QDI majority voter viz. SI_MV, CG_MV consumes 30.2% less area for physical realization, which makes it useful for resource-constrained fault tolerance applications such as those involving micro- or nano-systems used in small satellites where the hardware area is preferred to be a minimum.